\documentclass{aa}  

\usepackage{graphicx}
\usepackage{txfonts}
\usepackage{amssymb}
\usepackage{mathtools}
\usepackage{color}
\usepackage{siunitx}
\newcommand{\om}[1]{{\color{black} #1}}

\begin{document}

   \title{A spectroscopic study of MATLAS-2019  with MUSE: an ultra-diffuse galaxy with an excess of old globular clusters\thanks{Based on observations collected at the European Southern Observatory under ESO programme 0103.B-0635(B).}}
\titlerunning{A spectroscopic study of MATLAS-2019  with MUSE}
   \author{Oliver M\"uller\inst{1}, Francine R. Marleau\inst{2},  Pierre-Alain Duc\inst{1}, Rebecca Habas\inst{1},  J\'er\'emy Fensch\inst{3}, Eric Emsellem\inst{3,4}, M\'elina Poulain\inst{2}, Sungsoon Lim\inst{5},  Adriano Agnello\inst{6}, Patrick Durrell \inst{7}, Sanjaya Paudel\inst{8} , Rubén Sánchez-Janssen\inst{9}, Remco F. J. van der Burg\inst{4}}
   
\authorrunning{M\"uller, Marleau, Duc, Habas, et al.}
   \institute{Observatoire Astronomique de Strasbourg  (ObAS), Universite de Strasbourg - CNRS, UMR 7550 Strasbourg, France\\
 \email{oliver.muller@astro.unistra.fr} 
    \and
   Institut f{\"u}r Astro- und Teilchenphysik, Universit{\"a}t Innsbruck, Technikerstra{\ss}e 25/8, Innsbruck, A-6020, Austria   
   \and
Univ. Lyon, ENS de Lyon, Univ. Lyon 1, CNRS, Centre de Recherche Astrophysique de Lyon, UMR5574, F-69007 Lyon, France 
 \and
    European Southern Observatory, Karl-Schwarzschild-Str. 2, D-85748 Garching, Germany
\and
University of Tampa, 401 West Kennedy Boulevard, Tampa, FL 33606, USA
\and
DARK, Niels-Bohr Institute, Lyngbyvej 2, Copenhagen, Denmark
\and
Department of Physics \& Astronomy, Youngstown State University, Youngstown, OH 44555   USA
\and
Department of Astronomy, Yonsei University, Seoul 03722, Republic of Korea
      \and
   UK Astronomy Technology Centre, Royal Observatory, Blackford Hill, Edinburgh, EH9 3HJ, UK
 }

   \date{Received tba; accepted tba}

% \abstract{}{}{}{}{} 
% 5 {} token are mandatory
 
  \abstract
   {The MATLAS deep imaging survey has uncovered a plethora of dwarf galaxies in the low density environment it has mapped. A fraction of them are unusually extended and have a low-surface brightness. Among these so-called ultra-diffuse galaxies, a few seem to host an excess of globular clusters. With the integral-field unit spectrograph MUSE we have observed one of these galaxies --  MATLAS~J15052031+0148447 (MATLAS-2019) -- located towards the nearby group NGC~5846  and measured its systemic velocity, age, and metallicity, and that of its globular clusters candidates. For the stellar body of MATLAS-2019 we derive a metallicity of $-1.33^{+0.19}_{-0.01}$\,dex and an age of  $ 11.2^{+1.8}_{-0.8}$\,Gyr. For some of the individual GCs and the stacked GC population, we derive consistent ages and metallicities. 
   From the 11 confirmed globular clusters and using a Markov Chain Monte Carlo approach we derived a dynamical mass-to-light ratio of $4.2^{+ 8.6 }_{-3.4}$\,M$_\odot$/L$_\odot$. {This is at the lower end of the luminosity-mass scaling relation defined by the Local Group dwarf galaxies.} {Furthermore, we couldn't confirm nor reject the possibility of a rotational component of the GC system. If present,  this would further modify the inferred mass.} Follow-up observations of the globular cluster population and of the stellar body of the galaxy are needed to assess whether this galaxy is lacking dark matter like it was suggested for the pair of dwarf galaxies in the field of NGC\,1052, or if this is a miss-interpretation arising from systematic uncertainties of the  method commonly used for these systems and the large uncertainties of the individual globular cluster velocities.}

               
   \maketitle
%
%-------------------------------------------------------------------

\section{Introduction}
The Mass Assembly of early Type gaLAxies with their fine Structures (MATLAS) survey is a large observing program  designed to study low  surface  brightness  features  in  the  outskirts  of  nearby massive elliptical galaxies. The survey was conducted using MegaCam at the Canada-France-Hawaii Telescope (CFHT) and reaches surface brightnesses of $28.5-29.0$ mag/arcsec$^{-2}$ in the $g$-band while achieving  high image quality, enabling therefore the detection of low-surface brightness structures together with their globular cluster (GC) population  \citep{2015MNRAS.446..120D}. It is therefore an excellent data set to search for hitherto undetected dwarf galaxies. \citet{2020MNRAS.491.1901H} identified 2210 dwarf galaxy candidates with MATLAS. Among these, $\sim$ 4\% (Marleau et al., in prep.) fall into the category of the ultra-diffuse galaxies (UDGs). These are galaxies having effective radii larger than 1.5\,kpc and a low surface brightness \citep{1984AJ.....89..919S,2015ApJ...798L..45V} and appear in both cluster and field environments \citep{2016A&A...590A..20V}. Their extreme  low baryonic mass density makes them ideal probes for dark matter \citep{2019MNRAS.488L..24S,2019arXiv190901347S,2019MNRAS.484.4865P,2019ApJ...885..155W,2019ApJ...883L..33M} and alternative models of gravity \citep{2019MNRAS.487.2441H,2019A&A...627L...1B,2019PhRvD.100j4049I,2019MNRAS.482L...1M}.

Two of the most discussed UDGs (e.g. \citealt{2018MNRAS.481L..59H,kroupa2018does,2018ApJ...859L...5M,2019MNRAS.486.5670R,2019A&A...623A..36M,2019arXiv190708035N,2019MNRAS.489.2634H}) are the now famous NGC\,1052-DF2 \citep{2018Natur.555..629V} and NGC\,1052-DF4 \citep{2019ApJ...874L...5V}. These galaxies appear to have a deficiency of dark matter, based on their velocity dispersion measured from {a handful of} GCs these systems host  \citep{2018Natur.555..629V,2019ApJ...874L...5V}, and, in the case of NGC\,1052-DF2, the stellar body of the galaxy \citep{2019A&A...625A..76E,2019ApJ...874L..12D}. 
\om{If this interpretation holds, it would be a puzzling that an old dwarf galaxy with an age estimate of} 8.9$\pm$1.5\,Gyr \citep{2019A&A...625A..77F}
%and a rather metal poor stellar population of [Fe/H]=1.07$\pm$0.12\,dex \citep[as measured in ][]{2019A&A...625A..77F}, 
 hosts no massive dark matter halo. In the standard framework of cosmology a primordial dwarf galaxy  should always be surrounded a vast dark matter halo unless it is of tidal origin \citep{2012PASA...29..395K}.
 This interpretation of a lack of dark matter has been debated questioning the real distances of the galaxies \citep{2019MNRAS.486.1192T,2019ApJ...880L..11M,2019arXiv191007529D}, the effects of tidal interactions \citep{2018MNRAS.480L.106O,2019A&A...624L...6M}, or rotation \citep{2019A&A...625A..76E,2020MNRAS.491L...1L}.
The two UDGs NGC\,1052-DF2 and NGC\,1052-DF4, as the names suggests, reside in the same field. Therefore it is imperative to study such objects in different environments.

\begin{figure*}
    \centering
    \includegraphics[width=0.99\textwidth]{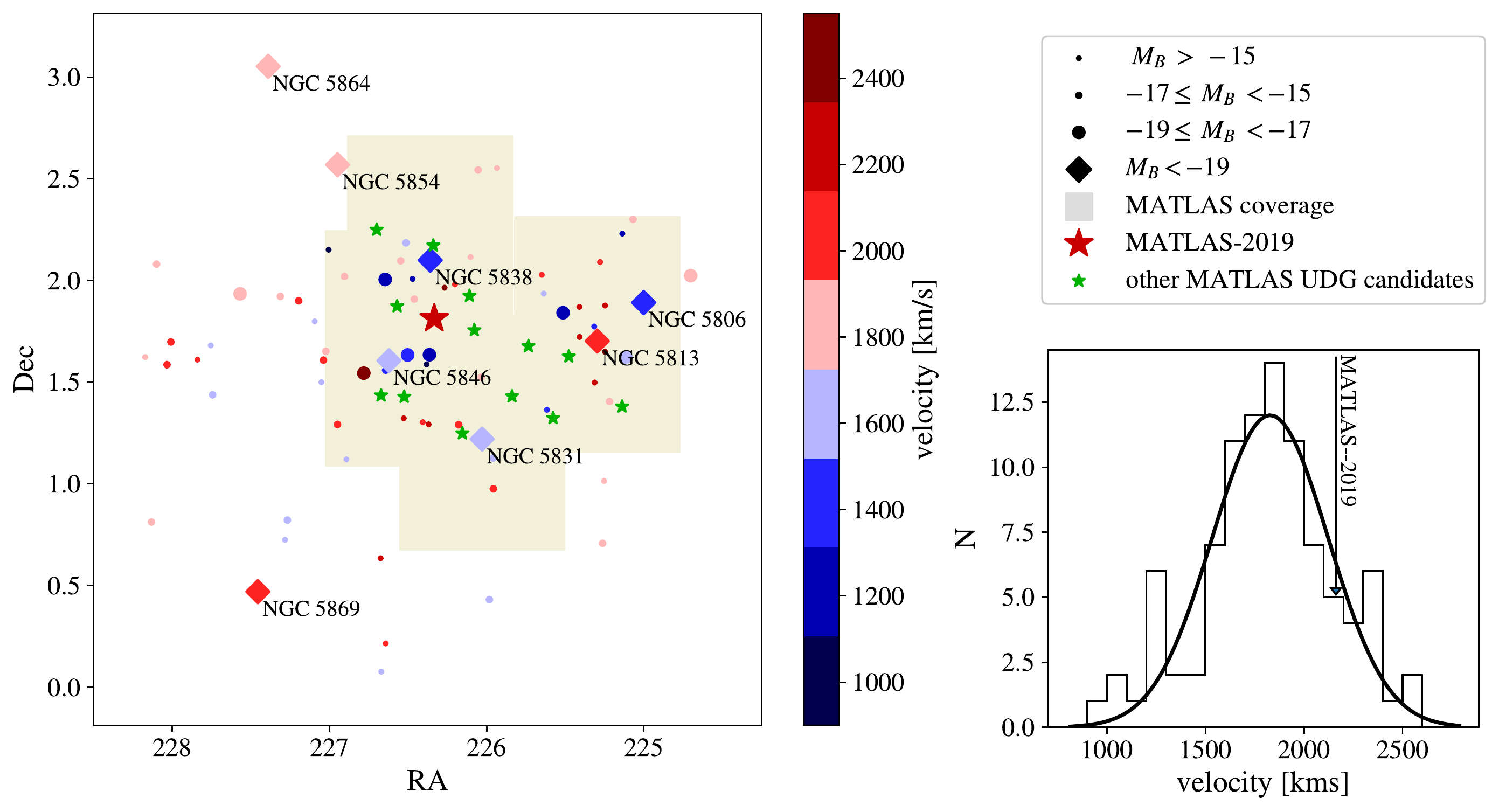}
    \caption{The environment of the NGC\,5846 group of galaxies. The large diamonds correspond to giant galaxies and the dots to dwarf galaxies, further dissected by their apparent magnitudes. MATLAS-2019 is indicated with the red star, the remaining UDG candidates are marked as green stars. The colors correspond to the velocities. The MATLAS survey field is indicated as shaded region. Velocities apart of MATLAS-2019 are taken from  \citet{2010A&A...511A..12E} {and presented as histogram in the bottom right corner}.}
    \label{env}
\end{figure*}

We have compiled a list of UDGs (Marleau et al., in prep.) from the MATLAS dwarf galaxy catalog \citep{2020MNRAS.491.1901H} and identified  a number of systems with a high number of globular clusters candidates  associated with them, reminiscent of NGC\,1052-DF2 and NGC\,1052-DF4. The GC catalog has been compiled from the MATLAS multi-band images, using color and size-proxy criteria to exclude foreground stars and background galaxies (see details in \citealt{2017ApJ...835..123L}).  
The candidate with the highest number of putative GCs -- MATLAS~J15052031+0148447, referred here as MATLAS-2019 (as this is the 2019th objects in the MATLAS dwarf catalog) -- was found in the field of the NGC\,5846 group of galaxies at a mean distance of 26\,Mpc \citep{2011MNRAS.413..813C}. 
This X-ray bright group is the most massive galaxy group in the nearby universe and has a mean velocity of $1828\pm295$ km/s, by considering the galaxies from \citet[][74 galaxies]{2010A&A...511A..12E}, \citet[][8 galaxies, after removing duplicates]{2015ApJS..217...27A}, \citet[][3 galaxies, after removing duplicates]{2005AJ....130.1502M}, and NED (9 galaxies, after removing duplicates). 
The central early-type galaxy NGC\,5846 is at a distance of 26.3\,Mpc, based on the GC luminosity function \citep{2009ApJ...690..512H}. MATLAS-2019 was identified as a potential group member by \citet{2005AJ....130.1502M}, and \citet{2019A&A...626A..66F} have independently conducted a detailed photometric study of this GC-rich UDG candidate based on data from the VEGAS survey. \om{At least in projection MATLAS-2019 is at the heart of the group (see Figure \ref{env}), with an on-sky separation of 0.35\,deg to NGC\,5846 (corresponding to 164\,kpc at the distance of  NGC\,5846).}

In this article, we present a spectroscopic analysis of the UDG candidate MATLAS-2019 and its rich GC population. Table\,\ref{table:galaxy} compiles all the important information on this galaxy. In Section\,\ref{reduction} we present the observations, data reduction and spectroscopy of the stellar body and the GCs, in  Section\,\ref{gcs_sec} we discuss the properties of the GCs, in Section\,\ref{darkmatter} we derive a dynamical mass estimate from the GC population and  discuss the amount of dark matter derived from the dynamical mass, and finally, in Section\,\ref{sum} we summarize our results.

\section{Observations, data reduction and spectroscopy}
\label{reduction}

\begin{figure*}
    \centering
    \includegraphics[width=0.32\textwidth]{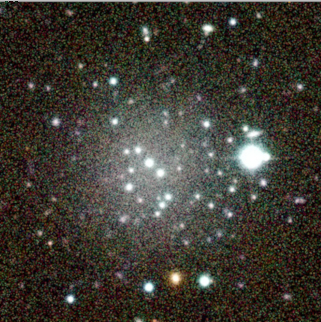}
    \includegraphics[width=0.32\textwidth]{MATLAS-2019-gri_annoted.png}
    \includegraphics[width=0.32\textwidth]{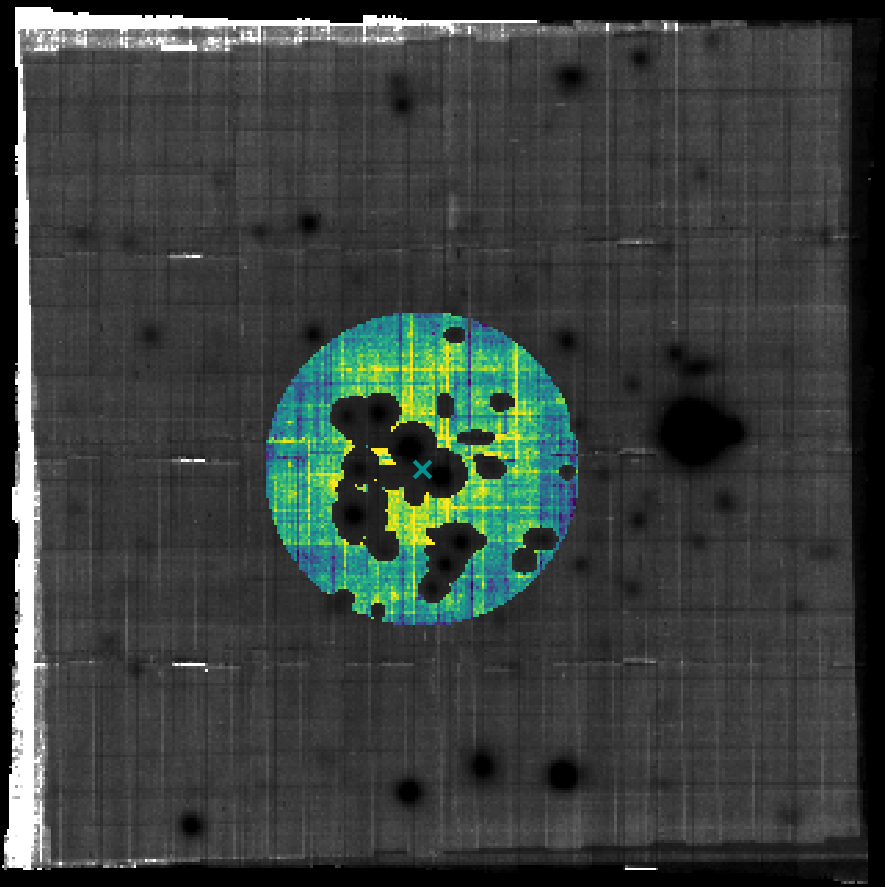}
    \caption{Left: The MATLAS true-color image (composite $g+r+i$). Middle: Residual g-band image, with a galaxy model subtracted.  The confirmed GCs (red) and GC candidates (blue) are labeled. The rejected MATLAS GC candidates are indicated in yellow.  Right: The region from which the galactic spectrum is extracted is indicated with the color map. Brighter colors correspond to a larger signal in the displayed MUSE stacked image. {North to the top, east to the left.} }
    \label{fig:image}
\end{figure*}

\begin{table}
\caption{Characterization of the UDG MATLAS-2019.}   
\label{table:galaxy}      
\centering       
\renewcommand{\arraystretch}{1.25}
\begin{tabular}{l l}     % 4 columns 
\hline\hline       
MATLAS-2019 & \\
\hline                    
RA &  15:05:20.2 \\
Dec &  +01:48:46 \\
assumed distance & 26.3 Mpc\\
m$_{\rm V}$ &$17.44\pm0.01$ mag\\
M$_{\rm V}$  & $-15.0$ mag\\
$\mu_{\rm eff,V}$  & 25.08 mag arcsec$^{-2}$\\
r$_{\rm eff,V} $ &  $17.2\pm0.2$ arcsec\\
 S\'ersic index $n$    & $0.73\pm0.01$ \\
 ellipticity &  $0.10 \pm 0.01$ \\
% $A_V$ [mag]& \\
v$_{\rm sys}$ & $2156.0\pm9.4$ km s$^{-1}$\\
r$_{\rm eff, {26.3\,\rm Mpc}} $ & $2187.6\pm25$ pc\\
$[$Fe/H$]$  & $-1.33^{+0.19}_{-0.01}$ dex\\
age &  $ 11.2^{+1.8}_{-0.8}$ Gyr\\
$\sigma_{\rm int}$ & $9.4^{+7.0}_{-5.4}$ km s$^{-1}$\\
L$_{\rm V}$  &  $8.59\times10^7$ L$_\odot$\\
 M${\rm_V}$/L${\rm_V}$  &   $ 2.0^{+0.3}_{-0.1}$ M$_\odot$/L$_\odot$\\
(M$_{\rm dyn}$/L$_\odot$)$_{\rm Wolf+2010}$  &$4.2^{+ 8.6 }_{-3.4}$ M$_\odot$/L$_\odot$\\
(M$_{\rm dyn}$/L$_\odot$)$_{\rm Errani+2018}$  &  $3.8^{+ 7.8 }_{-3.1}$ M$_\odot$/L$_\odot$\\

\hline                  
\end{tabular}
\end{table}

\begin{table*}
\caption{Positions and measurements of the unresolved/point-like sources and the UDG. }   
\label{table:sysvels}      
\centering          
\renewcommand{\arraystretch}{1.25}
\begin{tabular}{l l l l r l c c c}     % 4 columns 
\hline\hline       
Source & RA & Dec & m$_{\rm V}$ & S/N & v$_{\rm obs}$ & age & [Fe/H] & M${\rm_V}$/L${\rm_V}$\\
& [hh:mm:ss] & [dd:mm:ss] & mag & pix$^{-1}$ & [km/s] & Gyr & dex& M$_\odot$/L$_\odot$\\
\hline                    
UDG &  & &  & 12.4 &$ 2156.4\pm5.6$ & $ 11.2^{+1.8}_{-0.8}$ & $ -1.33^{+0.19}_{-0.01}$ & $ 2.0^{+0.3}_{-0.1}$\\
\\    
GC1 & 15:05:19.185 & $+$01:48:41.33 & 24.2  &3.3 & $ 2162.3\pm23.5$ & $ [7.2, 13.2] $ &  $ -1.26^{+0.60}_{-0.12}$ & $ [1.6, 2.6] $ \\ %
GC2 &  15:05:19.530 & $+$01:48:44.61 & 23.6 & 4.5 & $ 2138.5\pm23.3$ & $ [6.0, 13.7] $ &  $ -2.06^{+0.55}_{-0.21}$ & $ [1.2, 2.1] $  \\ % 
GC3 &  15:05:20.042 &$+$01:48:39.78& 23.5 & 5.9 & $ 2130.2\pm13.3$ &  $ 9.6^{+3.9}_{-0.7}$ &   $ -1.37^{+0.24}_{-0.24}$ & $ 1.7^{+0.6}_{-0.1}$ \\ %
GC4 &  15:05:20.122 &$+$01:48:38.20& 23.6& 5.5 &  $ 2133.6\pm17.2$ &  $ 9.1^{+4.9}_{-0.2}$ &  $ -1.22^{+0.20}_{-0.34}$ & $ 1.8^{+0.7}_{-0.1}$ \\ % 
GC5 & 15:05:20.141 & $+$01:48:44.61 & 22.9 & 11.1 & $ 2147.0\pm7.8$ &  $ 10.6^{+3.3}_{-1.4}$  & $ -1.32^{+0.17}_{-0.05}$ & $ 1.9^{+0.5}_{-0.1}$  \\ %
GC6 & 15:05:20.288 & $+$01:48:46.61 & 22.5 & 13.2 & $ 2147.2\pm5.0$ & $ 8.0^{+3.4}_{-0.3}$  & $ -1.26^{+0.07}_{-0.09}$  & $ 1.6^{+0.4}_{-0.1}$ \\ %
GC7 & 15:05:20.440 &$+$01:48:49.26 & 23.5 & 7.2 & $ 2157.2\pm13.8$ & $ 10.3^{+1.9}_{-3.7}$ & $ -1.76^{+0.27}_{-0.27}$ & $ 1.7^{+0.3}_{-0.5}$ \\ %
GC8 & 15:05:20.534 &$+$01:48:45.23& 24.4 & 4.5 & $ 2163.2\pm17.7$ & $ [7.6, 13.0] $ & $ -1.15^{+0.49}_{-0.01}$ &  $ [1.7, 2.7] $  \\ %
GC9 &  15:05:20.559 & $+$01:48:41.80 & 23.4 & 8.2 & $ 2179.1\pm13.7$ &  $ 11.5^{+1.9}_{-3.6}$ & $ -1.56^{+0.15}_{-0.30}$ & $ 1.9^{+0.2}_{-0.4}$ \\ %
GC10 & 15:05:20.593 &$+$01:48:48.87 & 24.3 & 3.7 & $ 2177.9\pm16.1$ & $ 11.3^{+1.6}_{-2.6}$ & $ [-1.5, -1.0] $ & $ [1.6, 2.3] $  \\ %
GC11 & 15:05:20.775 & $+$01:49:02.96 & 23.3 & 5.3 & $ 2134.2\pm18.9$ & $ [5.3, 12.2] $ & $ [-2.0, -1.4] $ & $ [1.1, 2.0] $  \\ %
\\
cand1 &  15:05:19.570 & +01:48:36.95 & 24.5 & &  \\ %
cand2 & 15:05:20.856 &+01:48:53.59 &23.8 & &  \\ %
\\
GC1-11   &  &  & & 19.4 & $ 2150.8\pm4.1$ & $ 9.1^{+3.0}_{-0.8}$  & $ -1.44^{+0.10}_{-0.07}$  & $ 1.6^{+0.3}_{-0.1}$ \\
cand1-2   &  &  & & 4.6 & $ 2184.0\pm12.8$ & $ 9.5^{+3.1}_{-3.5}$ &  $  -0.96^{+0.53}_{-0.15}$ & $ 1.9^{+0.7}_{-0.6}$\\
%GC &  RA & DEC & SNR & $\pm $ \\
\hline                  
\end{tabular}
\end{table*}

For the UDG candidate MATLAS-2019 we requested 12 OBs
with the Multi-unit spectroscopic explorer (MUSE) mounted at the Very Large Telescope at Cerro Paranal, of which 3 were taken in Period 103 (PI: Francine Marleau) under programme 0103.B-0635. 
\om{The data was reduced via the MUSE pipeline (Weilbacher et al., 2020) wrapped within the pymusepipe python package \textit{pymusepipe}\footnote{https://github.com/emsellem/pymusepipe}} \citep{2019A&A...625A..76E}, which was previously used to reduce MUSE data taken for the UDG NGC\,1052-DF2. This pipeline takes all raw data available in the ESO science archive and produces a combined and calibrated (i.e. bias and flat-field corrected, astrometrically calibrated, wavelength calibrated, and flux calibrated) data cube. The sky background was kept in the derived stacked data cube, and only then removed via the usage of ZAP the Zurich Atmosphere Purge (ZAP) packages \citep{2016MNRAS.458.3210S}, as it was done for NGC\,1052-DF2 \citep{2019A&A...625A..76E}. The total integration time on target was 7783\,s or 2.16\,hrs.

The extraction of the systemic velocity was done employing pPXF \citep{2004PASP..116..138C,2017MNRAS.466..798C} and follows the same procedure as we have employed for NGC\,1052-DF2 \citep{2019A&A...625A..76E,2019A&A...625A..77F}. \om{The stellar spectrum of the galaxy itself was extracted using a circular aperture (given its close to zero ellipticity): the radius of the aperture was set to 57\,px to optimise the S/N ratio.}
\om{To create a mask for the spectrum extraction, we have collapsed the full 3D cube into a 2D image. On this 2D image, }
point sources and background galaxies were masked based on the sources detected with Source Extractor \citep{1996A&AS..117..393B,2016JOSS....1...58B} with a $3\sigma$ threshold. Due to some blending issues we manually masked some of the brightest sources within the galaxy. Furthermore, we masked some additional sources which were not picked up by Source Extractor. After manually masking any strong sky features of the spectra, we can derive a systemic velocity with pPXF and the eMILES library \citep{2016MNRAS.463.3409V} using the most prominent absorption lines between 4800 and 8800\,\si{\angstrom}, namely H$\beta$, Mg, Fe, H$\alpha$, and CaT. However, depending on the signal-to-noise ratio (S/N) not all lines are visible. We use a set of Single Stellar Population (SSP) spectra with a Kroupa initial mass function (IMF), metallicities ([Fe/H]) ranging from solar down to -2.27 dex, and ages from 70 Myr to 14.0 Gyr. 
The  spectra from the SSP library were convolved with the line-spread function as described in \citet{2017A&A...608A...5G} (see also the Appendix of \citealt{2019A&A...625A..76E}).
A variance spectrum was measured on the masked data cube and added to pPXF. 
For the galaxy we measure a velocity of $2156.4\pm5.6$\,km/s. Because only 3 out of the 12 requested OBs were taken, the S/N is too low to apply a binning scheme for an estimation of the stellar velocity dispersion, as originally intended. To test if we can boost the signal, we have weighted the pixels  according to the S\'ersic profile of the galaxy with the optimal extraction algorithm \citep{1986PASP...98..609H} of MPDAF \citep{2016ascl.soft11003B}, giving more weight to the pixels near the center and less to the pixels in the outskirts, where fewer photons arrive. This, however, lead to a minimal change in the velocity estimation, due to the fact that the stellar profile is considerably flat -- one of the key properties of UDGs.

{The uncertainties of the velocities are derived via a wild bootstrap approach, as we have done in \citep{2019A&A...625A..76E}. Namely, at each wavelength we randomized the sign of the residual and added it to the best fit spectrum.} 
{We have repeated this 1000 times.} The $1\sigma$ standard deviation of the resulting velocity distribution then gives the error.

\begin{figure}
    \centering
    \includegraphics[width=0.49\textwidth]{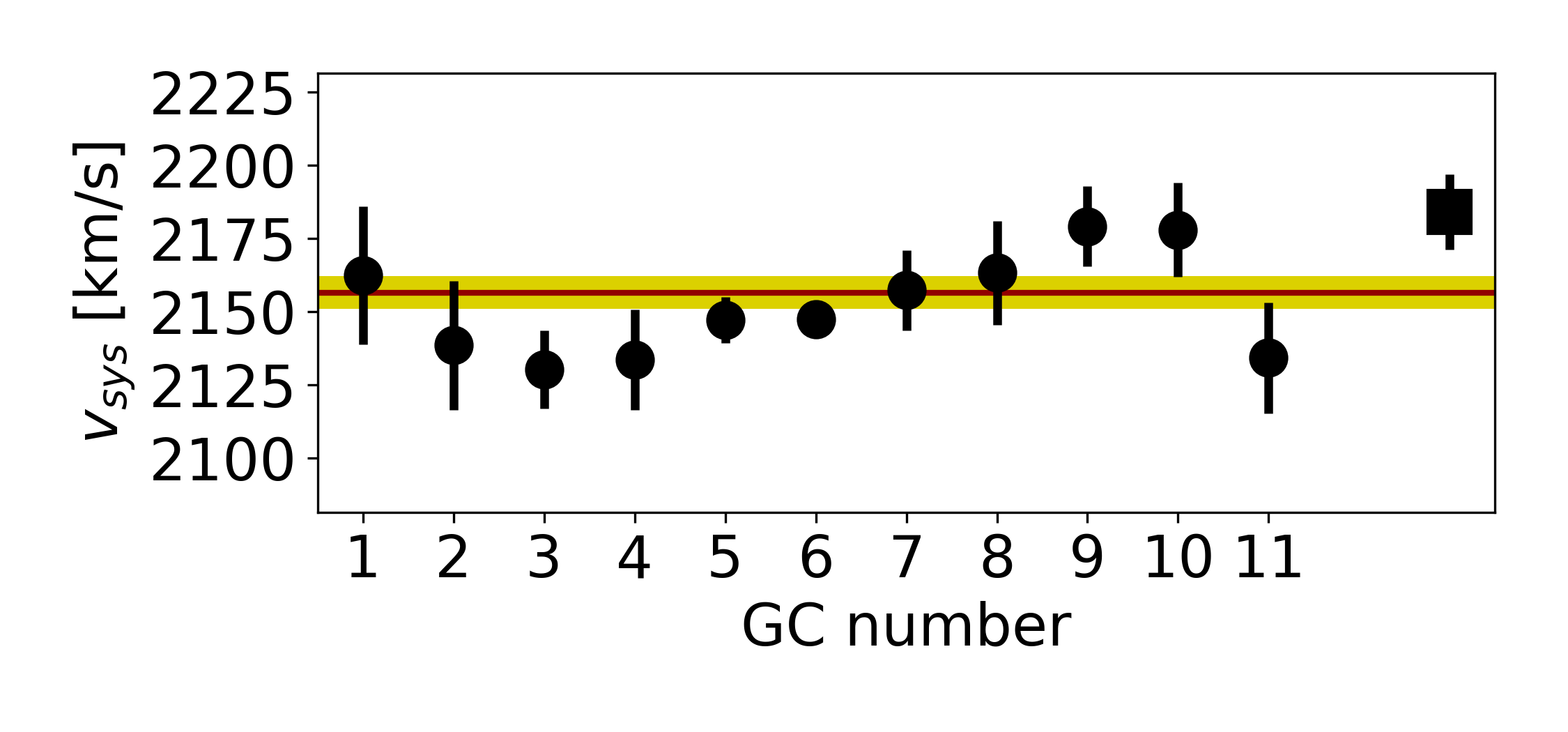}
    \caption{The velocities derived from the dominant absorption lines. The dots correspond to the velocities of the GCs, the square to the stacked spectra of the two remaining MATLAS GC candidates. The red line and shaded region indicate the velocity of the stellar body of the UDG and the corresponding uncertainty.}
    \label{fig:vels}
\end{figure}

To search for globular clusters in the MUSE data cube we have again run Source Extractor on the 2D image to find all point sources and have applied pPXF with circular apertures on top of the objects. To boost the signal, we have weighted the signal with a Gaussian with kernel width equals to the measured image quality ($\approx4.5$\,px or $1.1$\,arcsec). Point sources were rejected if the velocity was $\pm$100\,km/s away from the galactic velocity. \om{This range was selected to not miss any potential GCs with large uncertainties, but still being larger than the typical velocity dispersion of $\approx$20 to 30\,km/s for such low-surface brightness dwarf galaxies.}
Each remaining spectrum was then examined and a final GC catalog was produced. The S/N ratio per pixel is measured in a region between 6600 and 6800\,\si{\angstrom}. \om{It is calculated as the mean fraction between the flux and the square root of the variance. The variance itself was rescaled, i.e. multiplied by the Chi$^2$ value estimated from the best pPXF fit, to provide  a more direct account of the local noise.}
In total we find 11 GCs, {see Figure \ref{fig:image}. 10 out of the 11 GCs were in the GC candidate list based on the MATLAS $gri$ images. From this MATLAS GC list, two candidates have been uncovered as stars and another two are too faint for spectroscopy in MUSE. However, if we stack together these two candidates we get a reasonable spectrum on which we can derive a velocity ($2184.0\pm12.8$\,km/s), which is consistent with the velocity of the UDG. This indicates that these are likely also GCs of MATLAS-2019, so we consider them as GC candidates.} 
In Table\,\ref{table:sysvels} and  Figure\,\ref{fig:vels} we present the line-of-sight velocities for the galaxy, the GCs, and the stacked MATLAS GC candidates. 

\section{The GC system of MATLAS-2019}
In this section we discuss the properties of the GC system and its age and metallicity estimates.
\label{gcs_sec}

\subsection{\om{General properties of the GC system}}

\begin{figure}
    \centering
    \includegraphics[width=0.49\textwidth]{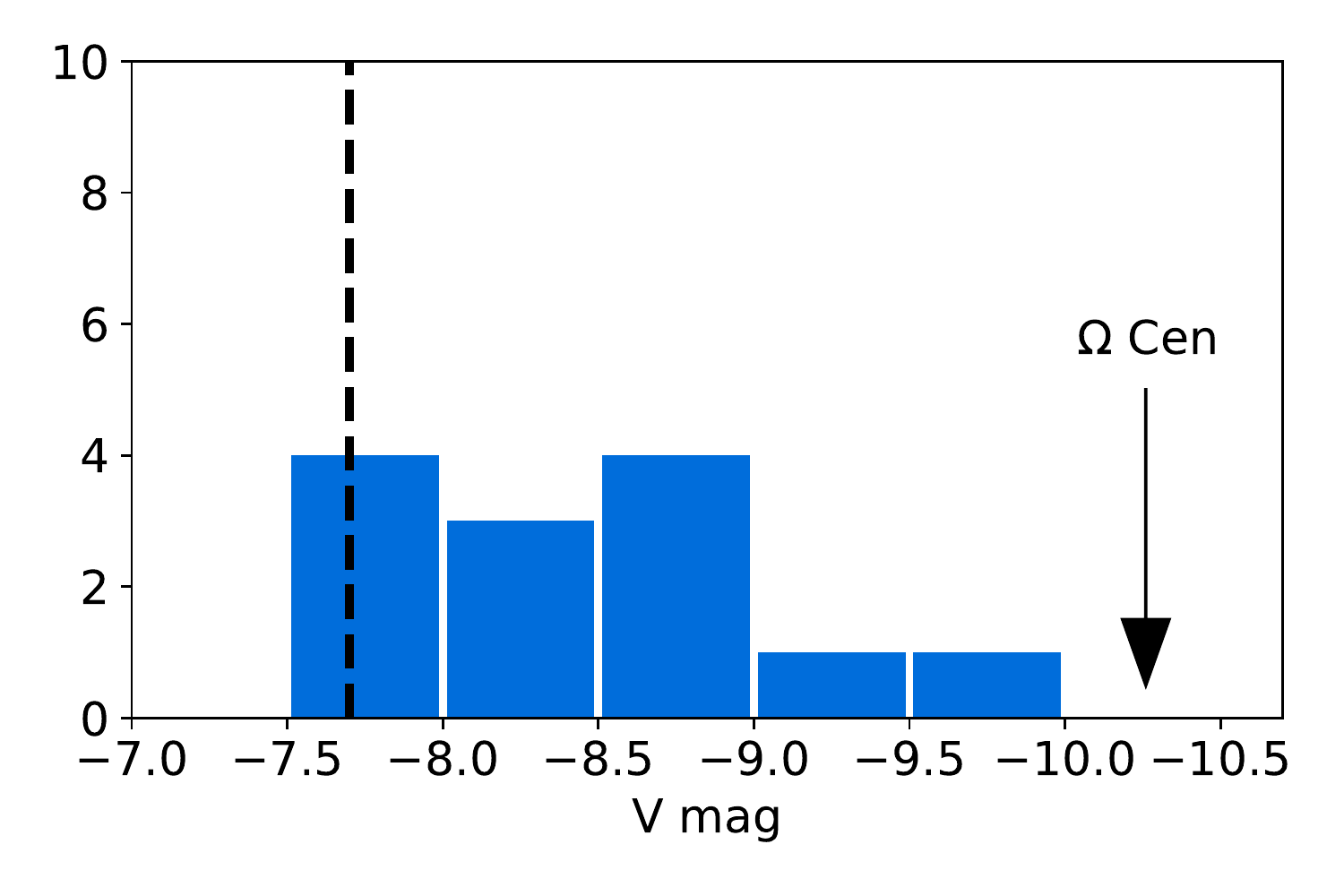}
    \caption{\om{The GCLF of MATLAS-2019 presented as a histogram, assuming a distance of 26.3\,Mpc. The black line denotes the putative peak of the GCLF at the given distance.}}
    \label{fig:gclf}
\end{figure}

\om{
The UDG MATLAS-2019 has a rich population of GCs. The GCs appear to be isotropically distributed, with no preferential alignment. The mean separation to the center of the galaxy is 8.0\,arcsec, which corresponds to 1.0\,kpc at the putative distance of 26.3\,Mpc. The most distant GC is at 19.6 arcsec within the MUSE field of view, i.e. a physical distance of 2.5\,kpc, which roughly coincides with the effective radius of the galaxy.  Half the GCs are within 6.0\,arcsec (0.8\,kpc). Interestingly, all the bright GCs are concentrated  in the central region of the galaxy (see Fig.\,\ref{fig:image}), with the brightest one, GC6, being located only  1.6\,arcsec (0.2\,kpc) away from the photometric galaxy center. The luminosity and putative distance of GC6 is $M_V=-9.6$ mag, making it compatible with both being a nuclear star cluster (NSC) and a GC 
(see e.g. Fig 8 in \citealt{2020A&A...634A..53F}). Its velocity is slightly offset from the velocity of the stellar body, but still well within the uncertainties. The total luminosity of GC6 is 19.6 mag in the $V$-band, which is $\approx$30 times fainter than that of the stellar body.
}

 \om{The brightest GC of MATLAS-2019 is with $M_V=-9.6$\,mag almost as bright as $\Omega$ Cen, see Figure \ref{fig:gclf}, which is indeed unexpectedly bright. This is similar to the} GC population of NGC\,1052-DF2 with the brightest GC having a similar luminosity as $\Omega$ Cen \citep{2018Natur.555..629V}. 
{The GC luminosity function \citep{2012Ap&SS.341..195R} and the specific frequency will be the topic of a future work.}

\subsection{Age and metallicity}

\begin{figure}
    \centering
    \includegraphics[width=0.49\textwidth]{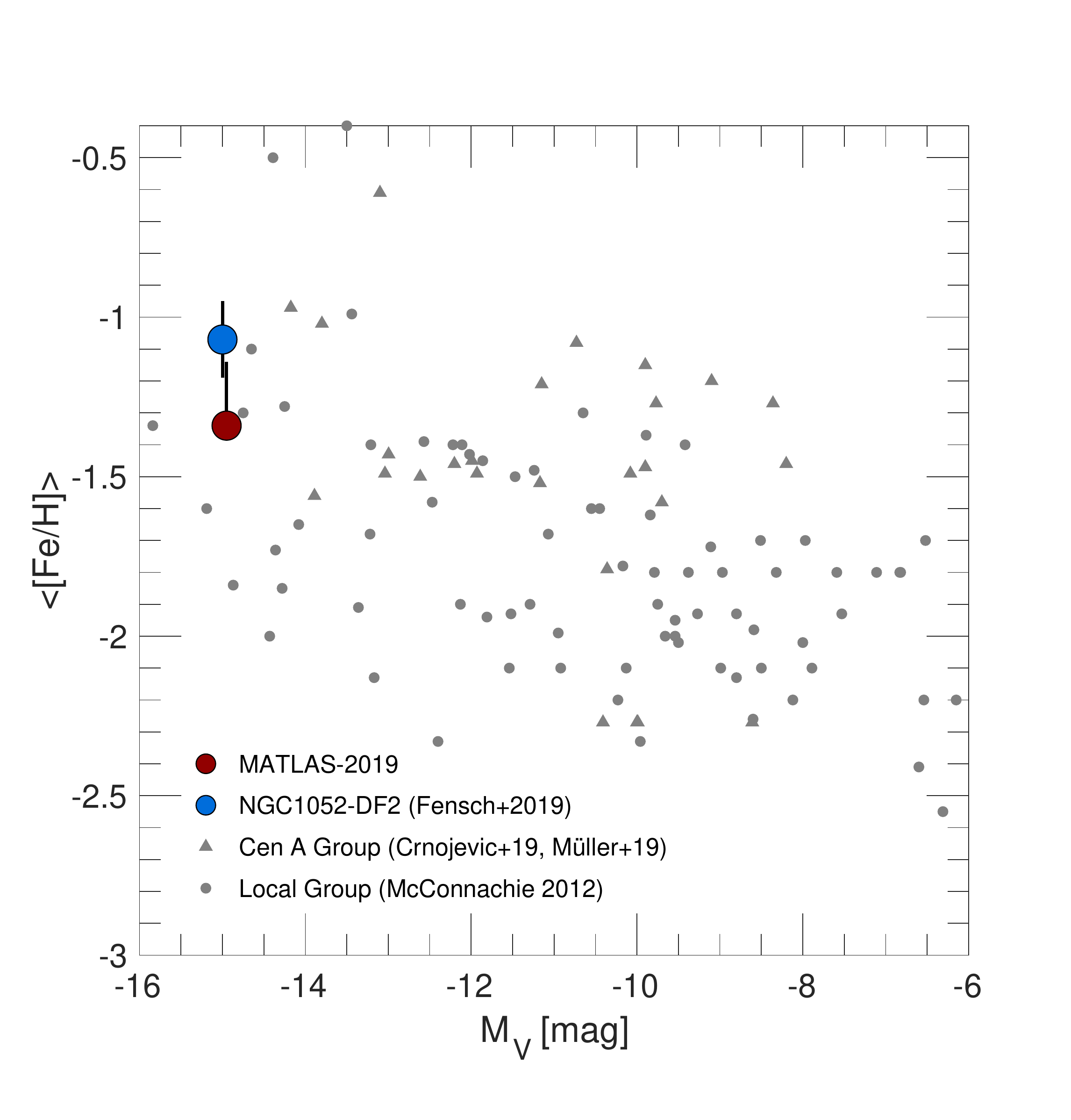}
    \vspace{-0.5cm}
        \includegraphics[width=0.49\textwidth]{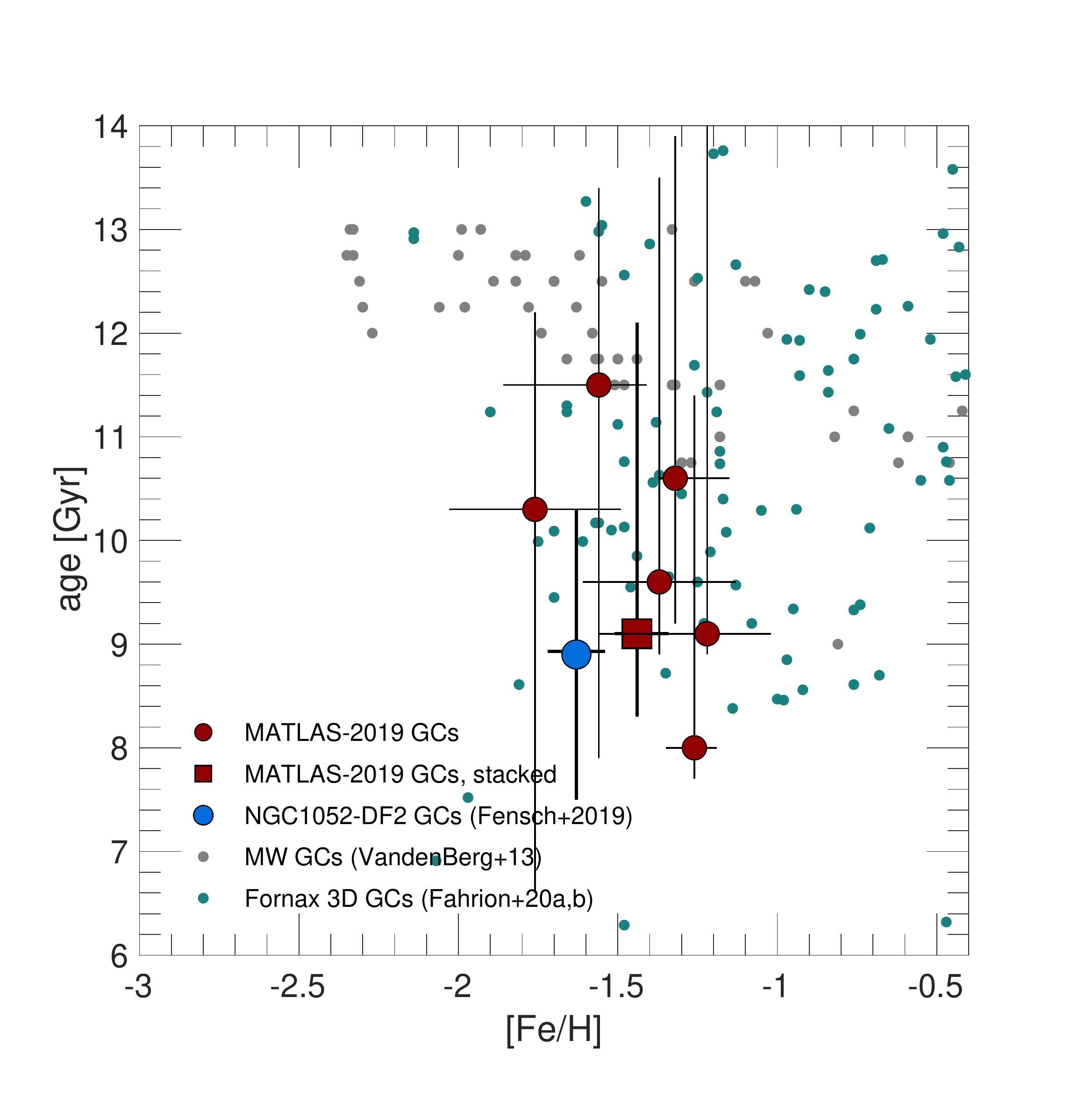}
    \caption{Top: The luminosity-metallicity relation for the Local Group dwarfs \citep[gray dots, ][]{2012AJ....144....4M}, Centaurus group dwarfs (gray triangles, \citealt{2019ApJ...872...80C,2019A&A...629A..18M}), NGC\,1052-DF2  \citep[blue point, ][]{2019A&A...625A..77F} and MATLAS-2019 (red point). Bottom: The metallicity-age relation for Milky Way GCs (gray dots,\citealt{2013ApJ...775..134V}), GCs from 23 galaxies of the Fornax cluster \citep[turquoise dots, ][]{2020arXiv200313705F,2020arXiv200313707F}, the stacked GC population of NGC\,1052-DF2  \citep[blue square, ][]{2019A&A...625A..77F}, and the GCs and stacked GC population of MATLAS-2019 (small red points and large red square). }
    \label{fig:relations}
\end{figure}

%Can we estimate metallicities and ages from the spectra? 
For some of the GCs {we were able to estimate a metallicity and age. We have derived these properties from the weights of the SSP models.} The estimations are provided in Table\,\ref{table:sysvels}.  Additionally, we have stacked the spectra of all the GC members of MATLAS-2019. With this we reach $S/N=19$ px$^{-1}$, which yields a more robust estimation of the mean metallicity and age of the GCs. For the stacked GC population, we derive a metallicity [Fe/H] $= -1.44^{+0.10}_{-0.07}$\,dex and an age of $9.1^{+3.0}_{-0.8}$\,Gyr. For the stellar body of MATLAS-2019 we derive [Fe/H] $= -1.33^{+0.19}_{-0.01}$\,dex and an age of $11.2^{+1.8}_{-0.8}$\,Gyr. These values are consistent with each other, a finding similar to the one in NGC\,1052-DF2 \citep{2019A&A...625A..77F}. The uncertainties are derived from the 16\% and 84\% percent intervals from the previous described bootstrap. {Where the bootstrap didn't converge, we present the interval in which 68\% of the estimates fall (again presented in Table\,\ref{table:sysvels}).} {From the weighted metallicities and ages we calculated the mass-to-light ratio of the stellar population. The uncertainties are again estimated from the bootstrap.}
In Figure \ref{fig:relations} we present how these numbers relate to the luminosity-metallicity relation of nearby dwarf galaxies and the metallicity-age relation for Milky Way GCs and GCs of massive galaxies of the Fornax cluster, as well as the properties derived for NGC\,1052-DF2 \citep{2019A&A...625A..77F}. \om{The metallicity of the MATLAS-2019 is consistent with other nearby dwarf galaxies and follows the luminosity-metallicity scaling relation. The GCs of MATLAS-2019 are also consistent with the metallicity-age relation as measured with MUSE in the Fornax 3D project \citep{2020arXiv200313705F,2020arXiv200313707F}. In this respect, there is nothing out of ordinary in these systems.}

\section{Dark matter content in MATLAS-2019}
\label{darkmatter}
\om{From the velocities of the GCs we can derive a velocity dispersion and ultimately the dynamical mass of the system. In the following we will infer the dynamical mass of MATLAS-2019 employing Bayesian considerations. We will first assume a completely pressure supported system and later generalize it for an additional rotational component.}

\subsection{Dynamical mass estimation of a pressure supported system}
\begin{figure}
    \centering
    \includegraphics[width=0.49\textwidth]{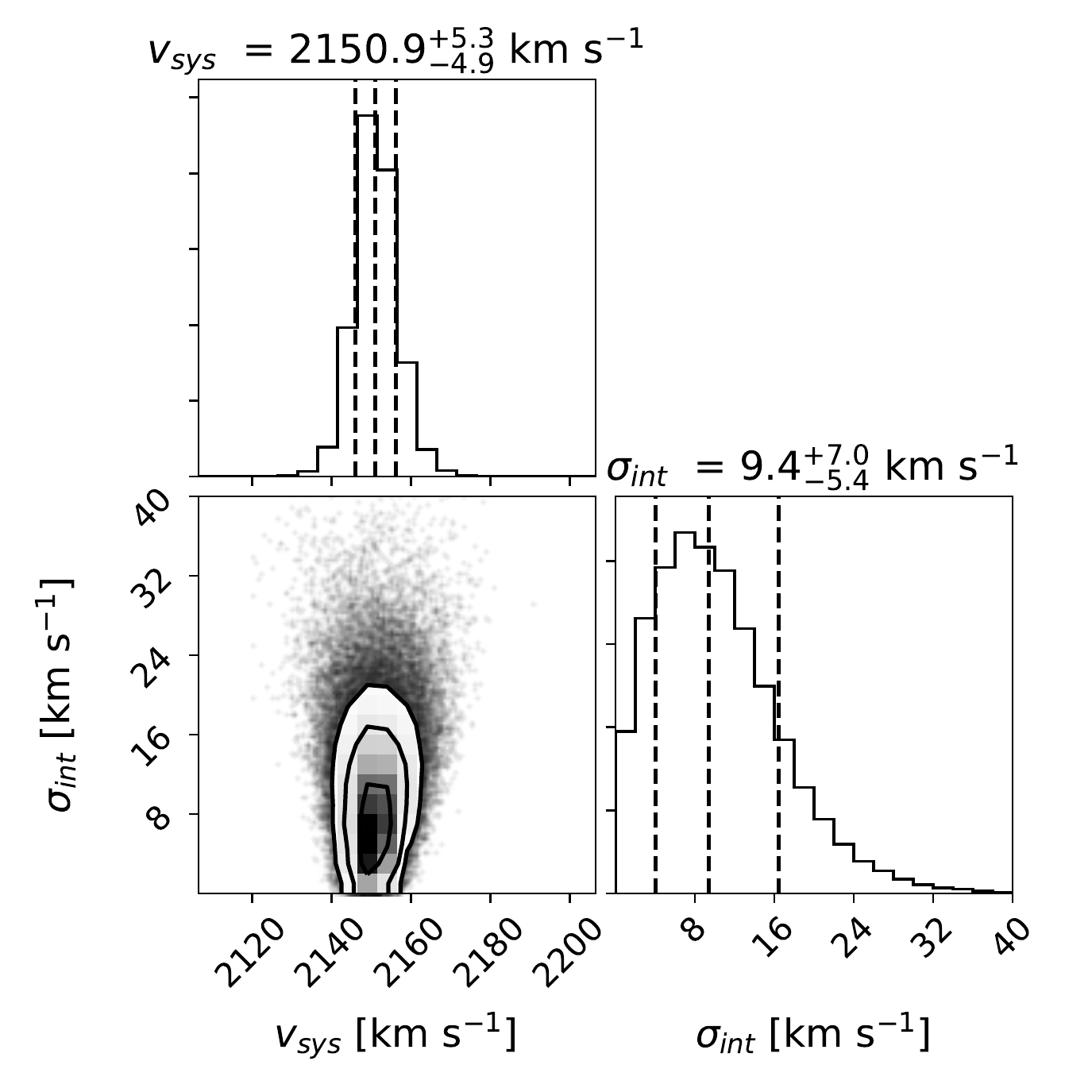}
    \caption{The sampled posterior distribution from our Markov Chain Monte Carlo analysis of the velocity dispersion and systemic velocity of the GC system. The three dashed lines indicate the 16, 50, and 84 percentiles, which correspond to the upper and lower uncertainty boundaries, and the best parameter estimation (i.e. the median).}
    \label{MCMC}
\end{figure}

\label{dynmas}
Assuming that the globular clusters are tracing the underlying gravitational potential\om{, are in dynamical equilibrium, and are pressure support dominated,} their velocity dispersion can be used to estimate the total mass of the system. \om{For this, we need to estimate the free parameters, namely the intrinsic velocity dispersion $\sigma_{\rm int}$ and the systemic velocity  $v_{\rm GCs}$ of all GCs combined.}
The log likelihood function is given by:
\begin{equation}
\log \mathcal{L}=\sum_{i=1}^N{\log \left(\frac{1}{\sqrt{2\pi} \sigma_{\rm obs}}\right) -\frac{(v_{{\rm obs},i}-v_{\rm GCs})^2}{2\sigma_{\rm obs}^2}},
\end{equation}
with
\begin{equation}
\sigma_{\rm obs}^2 =\sigma_{\rm int}^2+\delta_{v,i}^2,
\end{equation}
where $N$ is the number of tracers, $\sigma_{\rm obs}$ is the observed velocity dispersion, which is a combination of the true velocity dispersion $\sigma_{\rm int}$ and the observational uncertainties $\delta_{v}$, $v_{\rm obs}$ is the observed velocity, and $v_{\rm GCs}$ is the systemic velocity of all GCs combined. The two variables  $v_{\rm GCs}$ and $\sigma_{\rm int}$ are the parameters we are interested in. We impose a non-informative prior (Agnello \& Bruun in prep.), which suppresses too-small velocity dispersions\footnote{We have also tested a uniform prior of 1 for $v_{\rm GCs}$ between $\pm50$ around the mean of the observed velocities and $0<\sigma_{\rm int}<30$\,km/s.  Everywhere else the probability is set to 0. Using this prior instead will only slightly change the result on the order of 1\,km/s, which is well within the uncertainties. The difference is that the flat prior gives more realizations of very small velocity dispersions ($<$4\,km/s). These are highly unrealistic, as it would be less than what is expected arising from the baryonic content alone.}. We use a Markov Chain Monte Carlo (MCMC) approach  to sample over the two unknown parameters. For this purpose, we use the python package \textit{emcee} with 100 walkers, 100 iterations of burn-ins, and finally 10'000 steps along the chains. The resulting posterior distribution is shown in Fig.\,\ref{MCMC}. The errors are given by the 68 percent (i.e. 1$\sigma$ in frequentist statistics) bounds. We derive an intrinsic velocity dispersion of the GC and candidate system of  $\sigma_{\rm int}=9.4^{+7.0}_{-5.4}$\,km/s  and a systemic velocity of $v_{\rm GCs}=2150.9^{+5.3}_{-4.9}$\,km/s, respectively. 

Next, we estimate the dynamical mass-to-light ratio $M_{\rm dyn}/L_V$ within one de-projected half-light radius radius $r_{\rm 1/2}$. The dynamical mass  $M_{\rm dyn}$ within $r_{1/2}$ is given by \citep{2010MNRAS.406.1220W}
\begin{equation}
M_{\rm dyn}(r_{1/2})=\frac{4r_{\rm eff}\sigma_{\rm int}^2}{G}, {\,\rm with\,\,} r_{1/2}=\frac{4}{3}r_{\rm eff},
%M_{\rm dyn}(r_{1/2})=\frac{4r_{\rm eff}\sigma_{\rm int}^2}{G}, {\,\rm with\,\,} r_{1/2}=\frac{4}{3}r_{\rm eff}
\end{equation}

where G is the gravitational constant and $r_{\rm eff}$ is the measured effective radius (coming from a 2D S\'ersic fit). 
The luminosity $L_V$ is derived from the $g$-band magnitude using \citet{SloanConv} and a $(g-r)$ color of 0.59\,mag. With an absolute $V$-band magnitude of $-15.0$ (adopting a distance of 26.3\,Mpc) we get $L_V=8.59\times10^7$ L$_\odot$. The effective radius at a putative distance of 26.3\,Mpc is $r_{\rm eff}=2187.6$\,pc. Putting all together, this yields a dynamical mass within one de-projected half-light radius of  $M_{\rm dyn}=18.0^{+ 37.1 }_{-14.8}\times10^7$\,M$_\odot$ and finally a dynamical mass-to-light ratio of $M_{\rm dyn}/L_V=4.2^{+ 8.6 }_{-3.4}$\,M$_\odot$/L$_\odot$.
For the uncertainties in the distance, we have adopted a conservative lower limit of 22\,Mpc and a upper limit of 32\,Mpc, corresponding to the resp. lower and higher estimated distances of the massive galaxies in the field. 
{If we rather  use the more recent mass estimator by \citet{2018MNRAS.481.5073E}, {which updated the estimator provided by \citet{2011MNRAS.411.2118A}}, }
\begin{equation}
M_{\rm dyn}(1.8 r_{\rm eff})=\frac{6.5r_{\rm eff}\sigma_{\rm int}^2}{G},
\end{equation}
{
we derive a dynamical mass-to-light ratio within 1.8 $r_{\rm eff}$ (encompassing 87\% of the total
 luminosity) of  $3.8^{+ 7.8 }_{-3.1}$\,M$_\odot$/L$_\odot$. The two estimators yield consistent values, which shows that the choice of the mass estimator does not change the result.}

 How does this compare to other galaxies? To answer this question, we use the Spitzer Photometry \& Accurate Rotation Curves (SPARC) database provided by \citet{2016AJ....152..157L,2017ApJ...836..152L}, which gives a measure of the observed acceleration $g_{\rm obs}$ in terms of the acceleration expected by the baryons $g_{\rm bar}$. While this Radial Acceleration Relation (RAR) is strictly speaking purely observational, the deviation from unity gives information about the dark matter content of the galaxy. If $g_{\rm obs}$ is equals to $g_{\rm bar}$, the acceleration the galaxy experiences due to gravity can be solely explained by the baryonic content of the galaxy -- no need for dark matter. On the other hand, if $g_{\rm obs}$ is much larger than $g_{\rm bar}$ 
we need to invoke dark matter, or alternative gravity models \citep[e.g. MOND, ][]{1983ApJ...270..365M,2012LRR....15...10F} to explain the observations. 

\begin{figure*}
    \centering
    \includegraphics[width=0.49\textwidth]{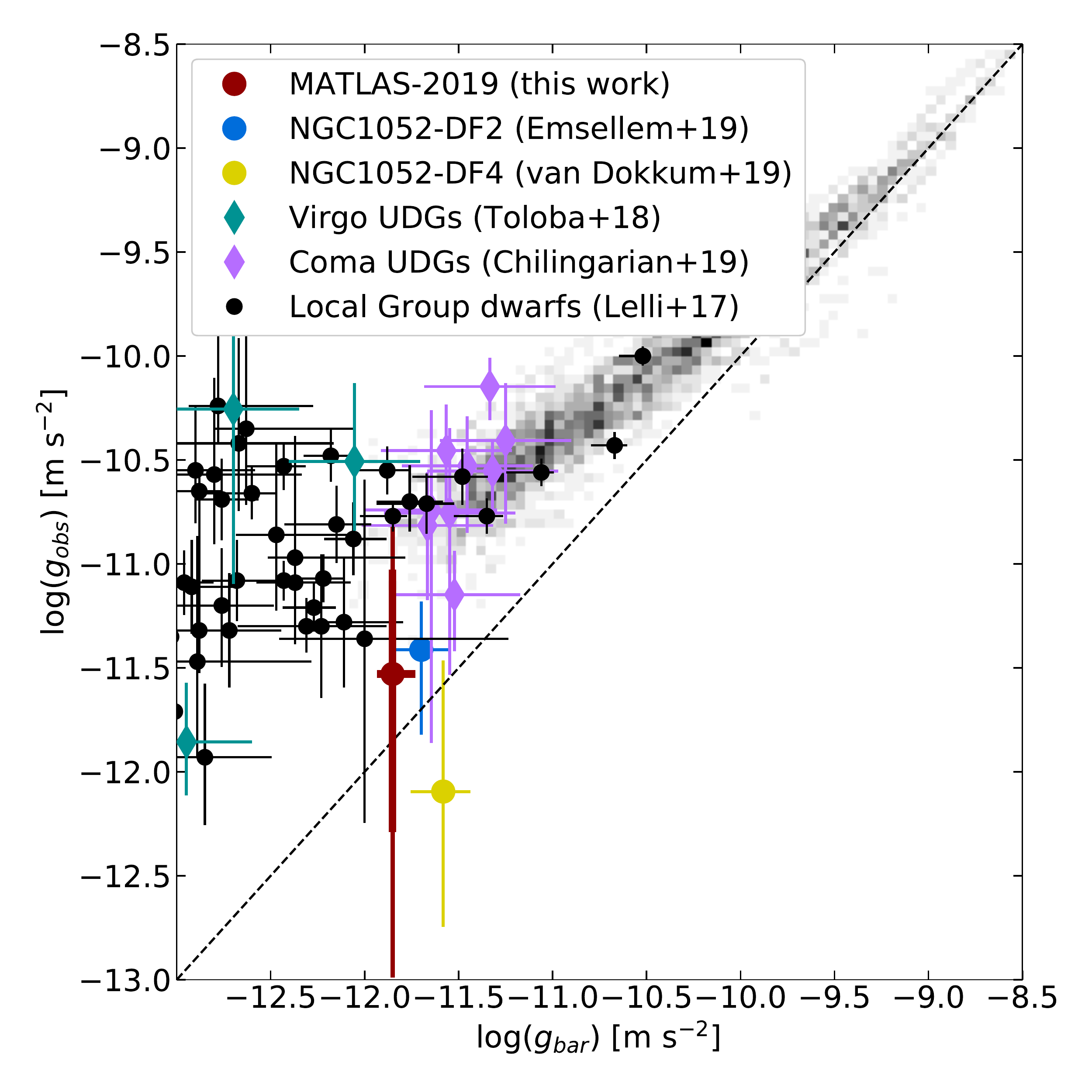}
    \includegraphics[width=0.49\textwidth]{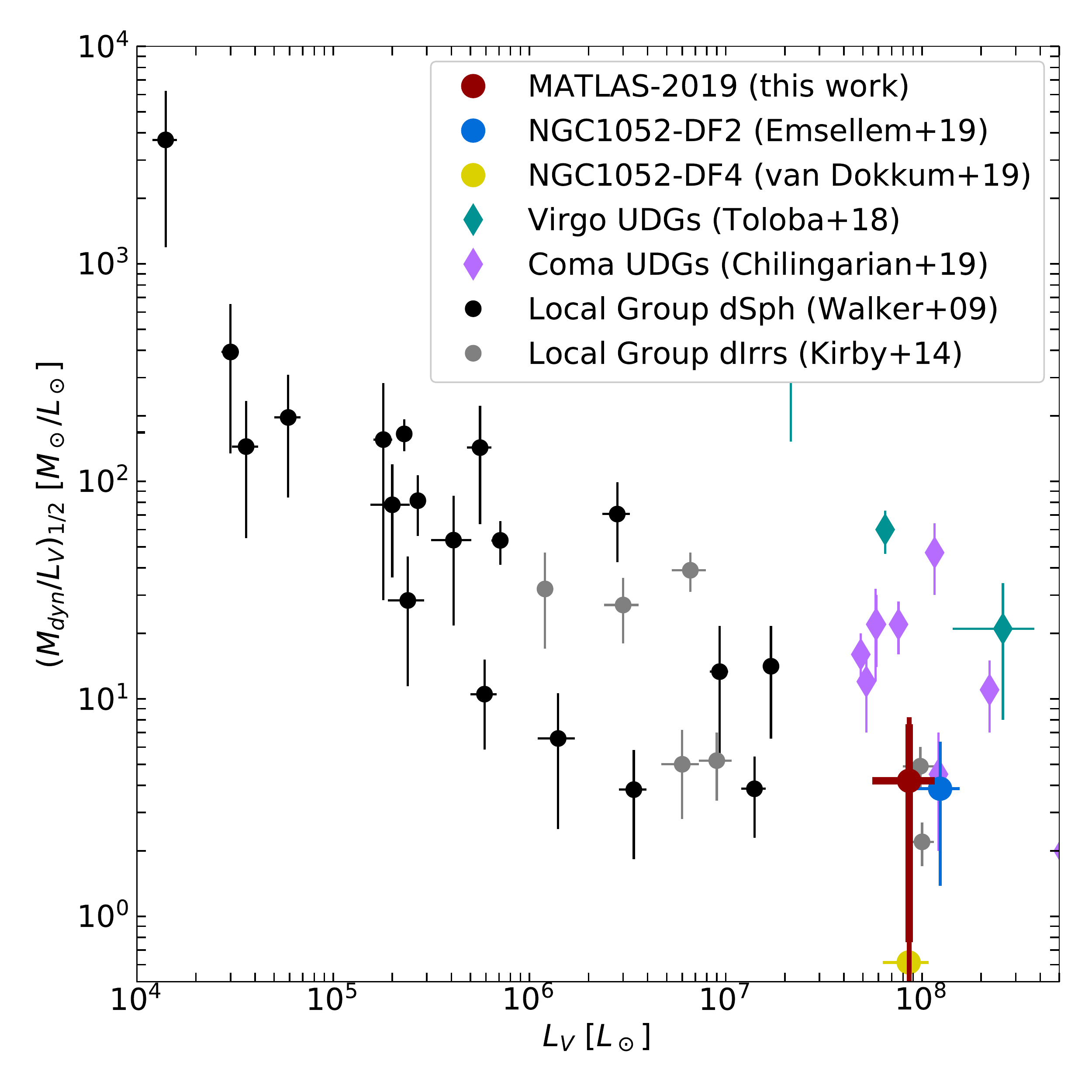}
    \caption{{ Left: The Radial Acceleration Relation (RAR) by \citet{2017ApJ...836..152L}. The black dots correspond to the Local Group dwarfs compiled in \citet{2017ApJ...836..152L}. The red dot plus lines give the measured values for MATLAS-2019 and its uncertainties (1 and 2 $\sigma$), respectively. The blue dot is the UDG NGC1052-DF2 observed with MUSE \citep{2019A&A...625A..76E}, and the yellow dot NGC1052-DF4 \citep{2019ApJ...874L...5V}. The cyan and violet squares are UDGs in the Virgo cluster \citep{2018ApJ...856L..31T} and the Coma cluster \citep{2019ApJ...884...79C}, respectively.
    The dashed line corresponds to unity.
    Right: The mass-to-light ratios for Local Group dwarf dSph (black dots, \citealt{2009ApJ...704.1274W}) and dIrr (gray dots, \citealt{2014MNRAS.439.1015K}), MATLAS-2019 (red dot), NGC\,1052-DF2 (blue dot, \citealt{2019A&A...625A..76E}), NGC\,1052-DF4 (yellow dot, \citealt{2019ApJ...874L...5V}), and the UDGs in the Virgo cluster (cyan diamonds, \citealt{2018ApJ...856L..31T})  and the Coma cluster (violet diamonds, \citealt{2019ApJ...884...79C}) as function of their luminosities.}
    }
    \label{fig:RAR}
\end{figure*}

From \citet{2017ApJ...836..152L} we can calculate:
\begin{equation}
g_{\rm obs}=\frac{3\,\sigma_{\rm int}^2}{r_{1/2}},
\end{equation}

\begin{equation}
g_{\rm bar}=\frac{\Gamma_V\,G\,L_V}{2\,r_{1/2}^2}.
\end{equation}
$\Gamma_V$ is the {stellar} mass-to-light ratio (M${\rm_V}$/L${\rm_V}$).
In Fig.\,\ref{fig:RAR} left panel, we show the RAR, together with our estimation for MATLAS-2019  the dwarf galaxies of the Local Group{, and other UDGs}. The observed acceleration of the UDG is close to unity, meaning that its baryonic content is  able to explain the measured velocity dispersion alone. With a $M_{\rm dyn}/L_V$ ratio of only 4, this is of course expected. Taking this at face value, the UDG appears to be lacking dark matter. However, the uncertainties are large. The {upper $1\sigma$ error bound} yields a $M_{\rm dyn}/L_V$ ratio of 14, which would be consistent with the RAR and the dwarf galaxies of the Local Group. {At $2\sigma$ the  $M_{\rm dyn}/L_V$ is 25, fully consistent with the Local Group dwarfs}. {This becomes even more evident when the  $M_{\rm dyn}/L_V$ ratio as a function of the luminosity is compared to Local Group dwarf galaxies. This is shown in Fig.\,\ref{fig:RAR}, right panel. The UDG, together with NGC\,1052-DF2 and NGC\,1052-DF4, follows the scaling relation as defined by Local Group dwarf galaxies and are comparable to the dwarf irregulars (dIrr) IC\,1613 and NGC\,6822 \citep{2014MNRAS.439.1015K}.}

\subsection{\om{Dynamical mass estimation with rotation}}
\begin{figure*}
    \centering
    \includegraphics[width=0.75\textwidth]{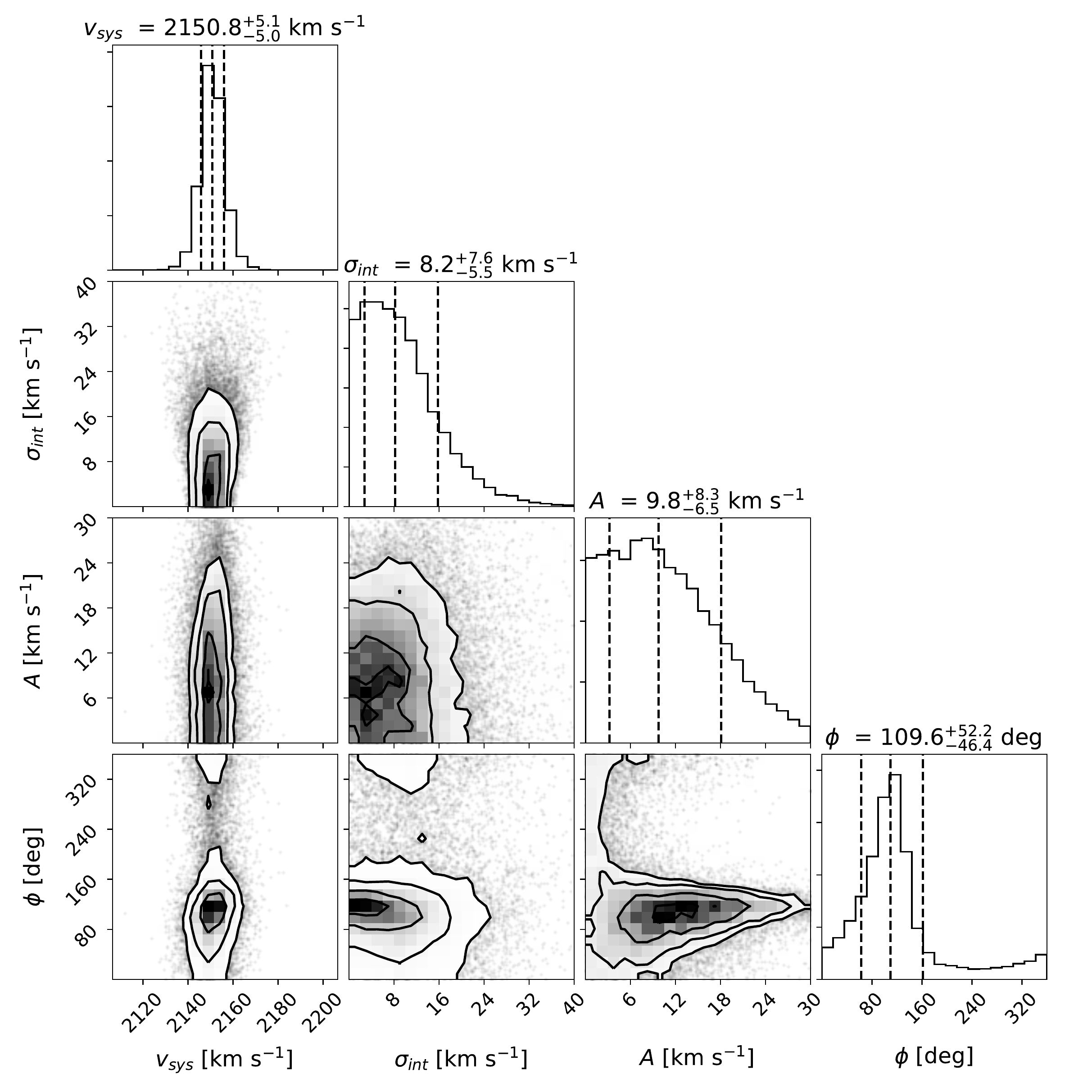}
    \caption{The sampled posterior distribution from our Markov Chain Monte Carlo analysis of the velocity dispersion and systemic velocity, including a rotational component of the GC system. The three dashed lines indicate the 16, 50, and 84 percentiles, which correspond to the upper and lower uncertainty boundaries, and the best parameter estimation (i.e. the median).}
    \label{MCMC_rot}
\end{figure*}

\begin{figure}
    \centering
    \includegraphics[width=0.49\textwidth]{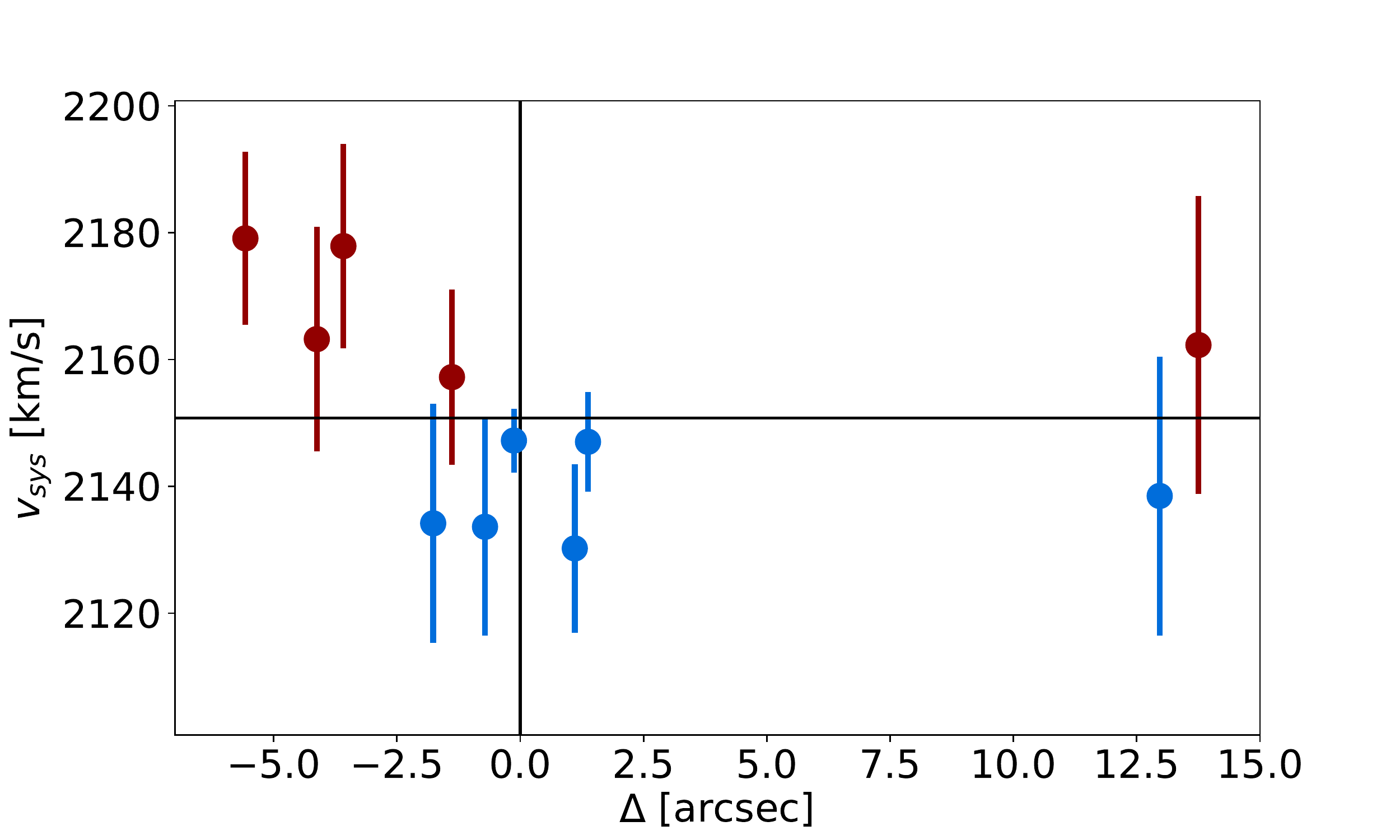}
    \caption{{The position-velocity diagram for the GC system, using the 2D distance to the best-fit rotation axis with angle $\theta$. Red and blue indicates whether the GC are red-or blueshifted in respect to the velocity of the GC system.}}
    \label{GC_rot}
\end{figure}

\om{
The previous assessment of the dynamical mass was based on the assumption that the system is fully pressure supported. An additional rotational component can change the results. In the case of NGC\,1052-DF2, a rotational signal for both the stellar body \citep{2019A&A...625A..76E} and the GC system \citep{2020MNRAS.491L...1L} was found, while the former could not be confirmed by independent measurements \citep{2019ApJ...874L..12D}. Let us now consider an additional rotational component for the GC system. For this, we will follow the description by \citet{2020MNRAS.491L...1L}.
The log likelihood function is given by:
\begin{equation}
\log \mathcal{L}=\sum_{i=1}^N{\log \left(\frac{1}{\sqrt{2\pi} \sigma_{\rm obs}}\right) -\frac{(v_{{\rm obs},i}-(v_{\rm rot}(\theta)+v_{\rm GCs}))^2}{2\sigma_{\rm obs}^2}},
\end{equation}
with
\begin{equation}
\sigma_{\rm obs}^2 =\sigma_{\rm int}^2+\delta_{v,i}^2,
\end{equation}
\begin{equation}
v_{\rm rot}(\theta) =A \sin{(\theta_i-\phi)},
\end{equation}
where $v_{\rm rot}$ describes the additional rotational component, $\phi$ the rotation axis, $\theta$ the angle between the line from the center of the galaxy to the GC and the east direction, measured counter-clockwise, and $A$ is the amplitude of the rotation velocity. We sample over the unknown parameters using a MCMC approach, as before. We use flat priors as it was done in \citet{2020MNRAS.491L...1L}. The results are shown in Figure\,\ref{MCMC_rot}. {We find a best separation of $\phi=110^{+52}_{-46}$ degrees and $A=9.8^{+8.3}_{-6.5}$ km s$^{-1}$. For the latter, the posterior distribution is mainly flat within 0 to 10 km s$^{-1}$. This posterior distribution and the best-parameter estimation with its errors cannot confirm nor rule out a rotational component of the GC system. In the following, we take the best-fit at face value and assume that there is indeed a rotational component for the sake of testing its impact on the mass estimation.}
{In Figure\,\ref{GC_rot} we present the position-velocity diagram for the best-fit rotation axis. For that, we have calculated the 2D separations of each GC to the rotation axis given by the angle $\theta$ and fixed at the center of the galaxy.}

Following the description of \citet{2020MNRAS.491L...1L} the dynamical mass is estimated with
\begin{equation}
M_{\rm dyn}(r_{1/2})= \Bigg(\Big( \frac{v_{\rm rot}}{\sin{(i)}}\Big)^2+ \sigma_{\rm int}^2 \Bigg)\, \frac{r_{\rm 1/2}}{G}.
\end{equation}
Here, an additional problem becomes evident -- we do not know the inclination $i$ of the rotational system {(if there is any)}. As the ellipticity of the UDG is close to zero, we can start by assuming that the inclination is 90 degrees, in other words, we see the rotation system perfectly edge-on. In this case, the $M_{\rm dyn}/L_V$ ratio is 2.6$^{+ 3.6 }_{-1.8}$. But in this case, the roundness of the object would be puzzling, as low-surface brightness dwarf galaxies tend to be best described as oblate-triaxial spheroids \citep{2019MNRAS.486L...1S}.
Is the shape of the dwarf galaxy a good indicator for the inclination of its GC system?  For NGC\,1052-DF2 this isn't the case. There, the rotation axis of the GCs was found to be roughly perpendicular to the major axis of the galaxy \citep{2020MNRAS.491L...1L}. If the GC system inherited a dynamic memory from an accretion event, no a priori alignment can be expected. Therefore, we cannot constrain the inclination with the ellipticity of MATLAS-2019.
Smaller inclinations will increase the $M_{\rm dyn}/L_V$ ratio to  3.0$^{+ 4.4 }_{-2.2}$ (60 degrees), 3.9$^{+ 6.0 }_{-2.8}$ (45 degrees), and 6.2$^{+ 11.1 }_{-4.6}$ (30 degrees). These values are again ambiguous, with the lower values indicating a lack of dark matter, and the larger values being consistent with dark matter dominated dwarf galaxies.
} {To conclude this, the data at hand cannot firmly constrain the existence of a rotational component, nor by including it what the mass of the galaxy would be.}

\subsection{MATLAS-2019 in Modified Newtonian Dynamics }
\label{MOND}
When the initial claim for the dark matter deficient UDG NGC\,1052-DF2 came up \citep{2018Natur.555..629V}, this was used as a falsification for alternative gravity models like Modified Newtonian Dynamics  (MOND, \citealt{1983ApJ...270..365M}, see also \citealt{2012LRR....15...10F}). In MOND-like theories the baryons invoke a dark matter behavior, so an absence of a phantom dark matter halo\footnote{in MOND, the term phantom dark matter  is used to describe a behavior as expected of dark matter in standard gravity. This means that the galaxy should exhibit a higher velocity dispersion than what is given by its baryonic content derived under Newton's law.} would be inconsistent with the theory. However, this assessment ignores a peculiar phenomenon in MOND, the so-called external-field effect (EFE, see e.g. \citealt{2019MNRAS.487.2441H} for a recent discussion), which can arise when a galaxy resides in an external gravitational potential. This EFE can lower the velocity dispersion of the system, making it appearing Newtonian, i.e. dark matter free. For NGC\,1052-DF2 it was shown that the EFE induced by NGC\,1052 can mitigate the tension \citep{kroupa2018does,2018MNRAS.480..473F}. 

So how does MATLAS-2019 fare in terms of MOND? In the isolated case, the expected MONDian velocity dispersion is calculated from the baryonic mass of the galaxy. Transforming its $V$-band magnitude with a M/L ratio of 2.0 this gives $\sigma_{\rm MOND}=17.9$\,km/s, which is marginally above the  one sigma upper limit of our measured velocity dispersion{, but well within two sigma.} 
The EFE calculation for the UDG cannot be conducted so easily, because the influence of all the nearby giant galaxies has to be taken into account. We refer to a future work but note that the EFE will push down the expected MONDian value of the velocity dispersion. One caveat though: if MATLAS-2019 is completely dominated by the EFE, it should be quickly dissolving, as the galaxy has no phantom dark matter protecting it against tidal forces (\citealt{2015MNRAS.454.3810M}, see also a similar discussion by \citealt{2019A&A...627L...1B} for the UDG DF-44 in the Coma cluster).

\section{Discussion and conclusions}
\label{sum}
With MUSE we have followed up the dwarf galaxy MATLAS 
J15052031+0148447 (MATLAS-2019) that has a rich globular 
cluster system and is located in the NGC\,5846 group of
galaxies. The object turned out to have been also detected in the VEGAS survey by \citet{2019A&A...626A..66F}.
 We got spectra of the stellar body and its GC candidates. We have confirmed 11 to be real GCs associated to the galaxy and  two additional as likely candidates. The mean velocity of the GC system derived from our MCMC approach is consistent with the velocity of the galaxy ($v_{\rm gal}=2156.4\pm5.6$\,km/s).  The velocity of the galaxy itself is consistent with the velocity distribution of the NGC\,5846 group of galaxies \om{($v_{\rm group}=1828.4\pm295.2$\,km/s).}
 \om{If the dwarf galaxy is at the distance of the NGC\,5846 group, its brightest GC would be rather intriguing, having a similar luminosity as $\Omega$ Cen.}
 
 For some of the GCs we were able to derive a metallicity and age. Additionally, the stacked GC spectrum allowed us to derive a metallicity and age estimate for the GC population and is with  [Fe/H] $= -1.44^{+0.10}_{-0.07}$\,dex and an age of $9.1^{+3.0}_{-0.8}$\,Gyr compatible with the one derived from the stellar body of MATLAS-2019 with [Fe/H] $= -1.33^{+0.19}_{-0.01}$\,dex and an age of $11.2 ^{+1.8}_{-0.8}$\,Gyr. This shows that both the galaxy and the GCs are old and metal-poor. Comparing the metallicities of the stellar body and the GCs to nearby dwarf galaxies and GCs, respectively, we find consistent results. 

From the individual GC velocities,  we have derived a velocity dispersion,  yielding $M_{\rm dyn}/L_V$ within one de-projected half-light radius of $4.2^{+ 8.6 }_{-3.4}$\,M$_\odot$/L$_\odot$. {Using another mass estimator within 1.8 times the effective radius, we derive a $M_{\rm dyn}/L_V$ ratio of 3.9$^{+ 8.1 }_{-3.1}$\,M$_\odot$/L$_\odot$, which is consistent with the previous estimate. Within the uncertainties, these values are consistent with the dark matter dominated dwarf galaxies in the Local Group, as well as the two apparently dark matter lacking galaxies in the NGC\,1052 group.}
In contrast to  NGC\,1052-DF2 and NGC\,1052-DF4, the association of MATLAS-2019, which belongs to a rich, X-ray luminous, group of galaxies should be much less ambiguous. 

\om{For the analysis of the dynamical mass, we have used the distance of 26.3\,Mpc of the central body of the galaxy group, namely NGC\,5846 ($v_{\rm NGC\,5846}=1712$\,km/s). There is a notably high difference in velocity between the two bodies. This could either mean that the UDG is on its infall into the group, or even farther behind. The latter would lower the the $M_{\rm dyn}/L_V$ ratio. In velocity space, the closest galaxies are NGC\,5869  ($v_{\rm NGC\,5869}=2065$\,km/s) and NGC\,5813 ($v_{\rm NGC\,5813}=1956$\,km/s). These galaxies have distance estimates of 24.9\,Mpc and 31.3\,Mpc, respectively. While the former would change the derived $M_{\rm dyn}/L_V$ ratio to 4.4\,M$_\odot$/L$_\odot$, the latter would lower it to 3.5\,M$_\odot$/L$_\odot$. The conclusions remain the same.}

{For NGC\,1052-DF2 indications of a rotational signal of the GC system was found. Could this be the case for MATLAS-2019 as well? Here,}  an analysis including the angles of the GCs for a rotational component of the GC system {remained inconclusive. For the moment, rotation cannot be ruled out.}  {Assuming that a rotation of the GC system is present,} the unknown inclination angle of the rotational component of MATLAS-2019 makes it difficult to confine the $M_{\rm dyn}/L_V$ ratio. An edge-on system would lead to a dark matter lacking galaxy, a larger inclination would rather make it consistent with dark matter dominated dwarf galaxies. {For the future, the confirmation of a rotational component could}   shed new light for the formation scenario of these systems. It is possible that the GCs were accreted and retained their dynamical memory, which could further lead to miss-interpretation of the mass content of these galaxies.

The GC population is not always a good indicator for the mass of a system, as it was shown for the Fornax dwarf spheroidal \citep{1991AJ....102..914M}. For Fornax, the GC system can yield a total mass of the object free of dark matter, while the stellar body suggests a $M/L$ ratio of 10 and more \citep{2009ApJ...704.1274W}, leaving ample space for dark matter. \citet{2019MNRAS.484..245L} further showed that even when solely considering the GC system of Fornax, the observation can  be interpreted as  ‘overmassive’,  ‘just  right’  or  ‘lacking dark matter’, due to large uncertainties from observations, mass estimators, scatter in the mass–concentration relation, and tidal stripping. A sample of just a few GCs yields order-of-magnitude \textit{systematic} uncertainties in the velocity dispersion and in the mass \citep{2019MNRAS.484..245L}, which will not be reflected in the given numbers presented here. Another caveat which has to be taken into account is that the system is not necessarily stationary enough to have reliable ensemble estimates, introducing even more biases. Even by ignoring these effects, the $1\sigma$ upper limit of the mass of the galaxy, the derived $M_{\rm dyn}/L_V$ ratio is consistent with other dwarf galaxies from the Local Group, therefore allowing for a fair share of dark matter. In other words, while the measured velocity dispersion of MATLAS-2019 taken at face value {could be interpreted as} a lack of dark matter, the uncertainties -- both systematic and observational -- do not rule out one or the other option. 
Therefore, measuring the stellar internal kinematics of the UDG is ultimately needed to understand whether there is a lack of dark matter in this galaxy.

\begin{acknowledgements}
{We thank the referee for the constructive report, which helped to clarify and improve the manuscript. The corner plots were created with the open source python package corner \citep{2016JOSS....1...24F}}
The authors thank Katja Fahrion for providing the table of GCs used in Figure\ref{fig:relations}. O.M. thanks Nicolas Martin for interesting discussions concerning the dynamical mass estimation.
O.M. is grateful to the Swiss National Science Foundation for financial support. S.P. acknowledges support from the New Researcher Program (No. 2019R1C1C1009600) through the National Research Foundation of Korea. A.A. was supported by a grant from VILLUM FONDEN (project number 16599). This project is partially funded by the Danish council for independent research under the project ``Fundamentals of Dark Matter Structures'', DFF--6108-00470.
\end{acknowledgements}

\bibliographystyle{aa}
\bibliography{aanda}

   \end{document}